\documentclass[prb,aps,floatfix,amsmath,onecolumn]{revtex4}
\usepackage[section]{placeins}

\usepackage{stmaryrd}
\usepackage{amsmath}
\usepackage{epsfig}
\usepackage{array}
\usepackage{verbatim}
\usepackage{hyperref}
\usepackage{amssymb}
\usepackage{multirow}
\usepackage{graphicx}
\usepackage[export]{adjustbox}

\usepackage{color}
\usepackage{ulem}

\usepackage[cspex,bbgreekl]{mathbbol}

\newcommand{\beg}{\begin{equation}}

\newcommand{\en}{\end{equation}}

\newcommand{\eref}[1]{Eq.~(\ref{#1})}

\renewcommand{\emph}{\textit}

\newcommand{\beq}{\begin{equation}}

\newcommand{\eeq}{\end{equation}}
\newcommand{\barray}{\begin{eqnarray}}
\newcommand{\earray}{\end{eqnarray}}

\usepackage{braket}

\begin{document}

\title{Ballistic transport in the classical Toda chain with harmonic pinning}

\author{Joel L. Lebowitz$^{1,2}$, Jasen A. Scaramazza$^2$}
\date{\today}							

\affiliation{$^1$ Department of Mathematics, Rutgers University, Piscataway, NJ 08854, USA\\ $^2$ Department of Physics and Astronomy, Rutgers University, Piscataway, NJ 08854, USA}

\begin{abstract}
We investigate, via numerical simulation, heat transport in the nonequilibrium stationary state (NESS) of the 1D classical Toda chain with an additional pinning potential, which destroys momentum conservation. The NESS is produced by coupling the system, via Langevin dynamics, to two reservoirs at different temperatures. To our surprise, we find that when the pinning is harmonic, the transport is ballistic. We also find that on a periodic ring with nonequilibrium initial conditions and no reservoirs, the energy current oscillates without decay. Lastly, Poincar\'e sections of the 3-body case indicate that for all tested initial conditions, the dynamics occur on a 3-dimensional manifold. These observations suggest that the $N$-body Toda chain with harmonic pinning may be integrable. Alternatively, and more likely, this would be an example of a nonintegrable system without momentum conservation for which the heat flux is ballistic - contrary to all current expectations.
\end{abstract}
 
\date{\today}
\maketitle
\section{Introduction}
The transport of thermal energy in Hamiltonian systems is a problem of great theoretical and practical interest [\onlinecite{Lepri,BLR}]. In its simplest form, one considers heat flow in the nonequilibrium stationary state (NESS) of a system in contact with two thermal reservoirs at different temperatures. Very little is known rigorously about this problem except in the case of harmonic crystals [\onlinecite{RLL}] or hard rods in 1D [\onlinecite{spohn}]. These models are special cases of the larger class of integrable models, whose extensive numbers of conserved quantities are expected in general to lead to ballistic heat transport [\onlinecite{zotos,mazur,suzuki,kudh}]. This means that if a system of length $N$ (and cross-section $A$) is put in contact with heat reservoirs at temperatures $T_L$ and $T_R$, $T_L > T_R$, at its left and right ends, then the heat flow in the stationary state $J$ would be (except for boundary effects) independent of $N$. This is what is observed for the Toda lattice and it stands in contrast to the case where we have dissipative transport satisfying Fourier's law, where $J$ would be proportional to $N^{-1}$.

In the absence of exact results, one has to rely on heuristics and simulations. A large number of these have focused on 1D systems. These have led to the following commonly accepted truths (CAT): Integrable systems such as the Toda chain [\onlinecite{shyo, toda2, zotos}], the Calogero-Moser system, the harmonic chain, and hard rods have ballistic transport, i.e., $J \sim N^{0}$. Nonlinear non-integrable systems such as the Fermi-Pasta-Ulam (FPU) chain or the diatomic Toda chain [\onlinecite{hatano}] have $J \sim N^{-\alpha}$ with $\alpha < 1$, for the momentum conserving case: the actual value of $\alpha$ depends on the system -- see [\onlinecite{waliha}] and references therein. When a nonintegrable system does not conserve momentum due to pinning by a one body potential, the transport is diffusive, also called ``normal'', with $\alpha = 1$.
	
We find, via numerical simulations, that the Toda chain with harmonic pinning has ballistic transport of heat. Because this system is generally believed to be nonintegrable, either the prevailing wisdom about 1D transport needs modification or this system is in fact integrable. In either case, the result is rather surprising and requires further investigation. We note that the Poincar\'e sections of the 3-body case indicate that the dynamics take place on a 3-dimensional manifold for all tested initial conditions, indicating that there are 3 conserved quantities in this case (the first two being the Hamiltonian itself and a quantity corresponding to the harmonic motion of the center of mass). We also note that when the pinning is done by a quartic potential, then the heat transport is clearly not ballistic, although we cannot give clear evidence that $\alpha = 1$. We suspect the transport is indeed diffusive, as quartic pinning is sufficient to induce diffusive scaling in the harmonic chain [\onlinecite{aolusp}].

\section{The model}
Consider a 1-dimensional chain of $N+2$ labeled particles, i.e., located on the lattice $L = \{0, 1, ..., N, N+1\}$, with the following classical Hamiltonian $H$
\beg
H = \sum_{i=0}^{N+1}\bigg[\frac{p_i^2}{2} + \frac{\nu^2}{z}q_i^z + V(r_i)\bigg],\quad r_i \equiv q_{i+1} - q_i,\quad z \textrm{ even}.
\label{Ham}
\en
Here $\{q_i\}$ are the displacements of the particles, $\{p_i\}$ are their momenta, $\nu$ is the strength of the one-body pinning potential and $V_i \equiv V(r_i)$ is the interaction potential. For periodic boundary conditions $V_{N+1} = V(q_0 - q_{N+1})$ while for fixed boundary conditions $V_{N+1} = 0$ and $q_0 = q_{N+1} = p_0 = p_{N+1} = 0$. When $\nu = 0$ and
\beg
V_i = \frac{a}{b}\exp[-b r_i],\quad a, b > 0,
\label{TodaPot}
\en
the system is the Toda chain [\onlinecite{toda}], which is a well-known integrable model for both periodic and fixed boundary conditions [\onlinecite{henon,flaschka}]. Unless otherwise specified, $V_i$ will refer to the Toda interaction for the remainder of this work.

In this note, we numerically investigate the heat transport properties of the fixed boundary Toda chain with the addition of an on-site harmonic potential, i.e., $\nu \ne 0$ and $z = 2$. In general, such a modification is expected to break the integrability of the $\nu = 0$ system when the number of particles is greater than 2. Indeed, the only obvious conserved quantities when $\nu \ne 0$ are $H$ itself and the center of mass term $h_{c}$
\beg
h_{c} = \frac{1}{2}\bigg(\sum_{i=0}^{N+1} p_i\bigg)^2 + \frac{\nu^2}{2}\bigg(\sum_{i=0}^{N+1} q_i \bigg)^2.
\label{c}
\en
We couple particles 1 and $N$ of the chain to Langevin baths with a coupling constant $\mu$, which act as thermal reservoirs at temperatures $T_L$ and $T_R$ and induce a nonequilibrium steady state (NESS). The infinitesimal generator of motion $\mathcal{L}$ is therefore
\beg
\begin{split}
\mathcal{L}(\cdot) &= \mu B_{1,T_L}(\cdot) + \mu B_{N,T_R}(\cdot) + \mathcal{A}(\cdot),\\
A(\cdot) &=\sum_{j=1}^{N}(p_{j+1}-p_j)\partial_{r_j} + \sum_{j=1}^{N}(V'_j-V'_{j-1} - \nu^2 q_j^{z-1})\partial_{p_j},\\
B_{j,T_{L/R}}(\cdot) &= - p_j\,\partial_{p_j} + T\,\partial^2_{p_j},\quad j = 1,\,N;\quad \mu = \textrm{ bath coupling}
\end{split}
\label{TotalGen}
\en
where $V'_j \equiv \frac{dV(r_j)}{dr_j}$. For systems like \eref{TotalGen}, the integrability of the bulk dynamics plays a central role in determining the transport properties [\onlinecite{zotos}], although in the quantum mechanical case this statement requires qualification [\onlinecite{wube}]. The central quantity of interest is the average heat current $J$, which in the NESS is given by
\beg
\begin{split}
J = \langle J_j \rangle = -\bigg\langle\frac{1}{2}(p_j+p_{j+1})V'_{j}\bigg\rangle,\quad j \in [2,N-1],
\end{split}
\label{NESSJ}
\en
where $\langle\cdot\rangle$ refers to the NESS average, which in simulations is computed by first allowing the system sufficient time to relax to the NESS before time averaging. Also of interest is the NESS temperature profile $T_j$
\beg
\begin{split}
T_j = \langle p_j^2 \rangle,\quad j \in [1,N].
\end{split}
\label{NESSTj}
\en
In the following, we will give evidence that when $z=2$, $J \sim N^0$ and that $T_j$ is independent of $j$ in the bulk, with a jump in $T_j$ at the reservoirs. These two properties are only expected to hold when the bulk dynamics are integrable. Indeed, because these bulk dynamics are expected to be nonintegrable and break translational invariance, one would expect $J \sim N^{-1}$ and $T_j$ to be a continous curve interpolating between $T_L$ and $T_R$. We then show that when the pinning is anharmonic, e.g., $z = 4$, then the system satisfies the ordinary expectations. Spurred by the unexpected harmonic pinning result, we then give further evidence of nondissipative behavior in this model.

\section{Nondissipative behavior}
\subsection{Ballistic transport}
\label{btransport}
\begin{figure}
\includegraphics[width=.8\linewidth]{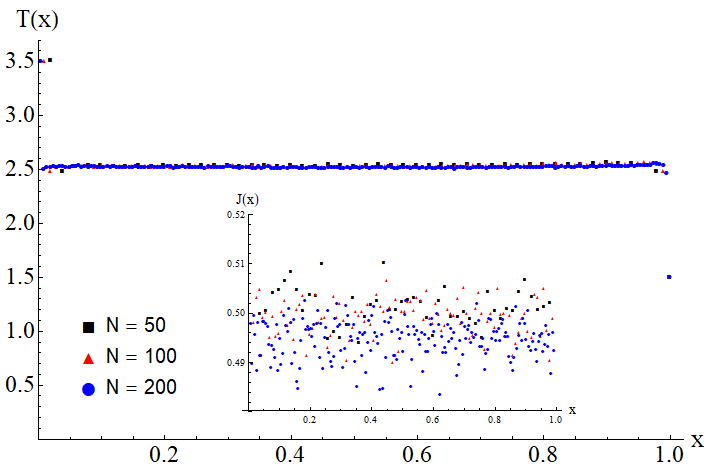}
\caption{Temperature profiles ($T_L = 4$, $T_R = 1$) for the Toda chain with harmonic pinning ($a = b = \mu = \nu = 1$, $z = 2$). The $j$-th particle occupies position $x_j = j/N$. Each profile is an average over 6 runs. The dynamics were integrated with a timestep $dt = 0.005$. We allow $2\times10^9$ timesteps to reach the NESS and then average the currents and temperature profiles over the same amount of time, with $2\times10^4$ timesteps in between each measurement. Insert: Averaged site-by-site current profile. The fact these profiles are flat indicates that we averaged in the NESS.}
\label{harmpin_N}
\end{figure}

\begin{figure}[!h]
\includegraphics[width=.8\linewidth]{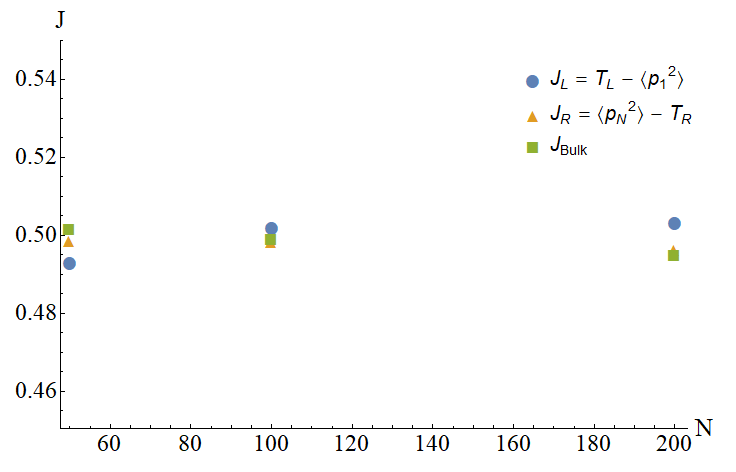}
\caption{The currents of the temperature profiles from Fig.~\ref{harmpin_N} as a function of $N$. There is no appreciable decay of the currents as $N$ increases, i.e., the transport is ballistic. The ratios of the bulk currents are $J_{100}/J_{50} = 0.995$ and $J_{200}/J_{100} = 0.992$.}
\label{harmpin_N_2}
\end{figure}

\begin{figure}[!h]
\includegraphics[width=.8\linewidth]{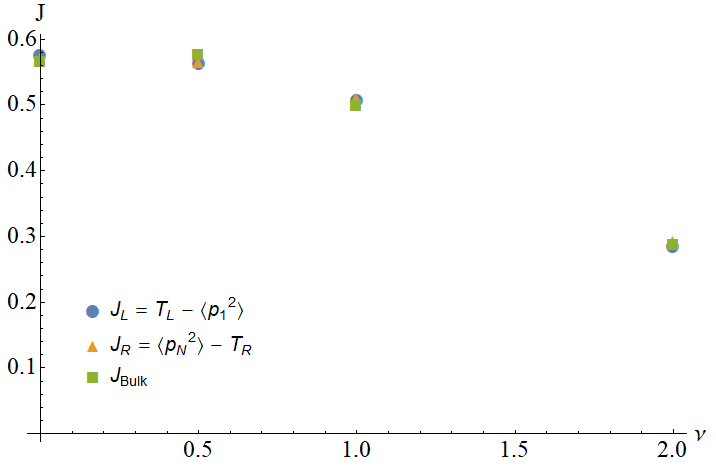}
\caption{Observed decrease in current as the pinning strength $\nu$ is increased for the driven Toda chain with harmonic pinning. Here $N = 100$ and all parameters are the same as in Fig.~\ref{harmpin_N}, save for the varying $\nu$ and a shorter relaxation and observation time, both $2\times10^8$ time steps. The three methods of measuring the current from \eref{Js} significantly overlap.}
\label{pinstrength}
\end{figure}

Consider the dynamics \eref{TotalGen} for the Toda interaction \eref{TodaPot} and $z=2$ (harmonic pinning). We integrate \eref{TotalGen} using a velocity Verlet algorithm adapted to include the Langevin reservoirs [\onlinecite{AllTil}]. In Fig.~\ref{harmpin_N}, we show the steady state temperature profiles for several $N$ and in Fig.~\ref{harmpin_N_2} we show the values of the corresponding steady state currents, which we measure in three ways. $J_{\textrm{Bulk}}$ is obtained by averaging $J$ from \eref{NESSJ} over the entire bulk of the chain, while $J_L$ and $J_R$ follow from the steady-state average of the energy flux from a Langevin bath
\beg
\begin{split}
J_{\textrm{Bulk}} = \frac{1}{N-2}\sum_{j=2}^{N-1}\langle J_j \rangle,\\
J_L = \mu(T_L - \langle p_1^2\rangle),\\
J_R = \mu(\langle p_N^2 \rangle- T_R).
\end{split}
\label{Js}
\en
The agreement of the three numerically obtained currents from \eref{Js} indicates that we have indeed reached the steady state.
The flat temperature profile and non-scaling current illustrated in Fig.~\ref{harmpin_N} and Fig.~\ref{harmpin_N_2} is expected for the $\nu = 0$ (integrable) case, but it is entirely surprising for $\nu \ne 0$. In Fig.~\ref{pinstrength}, we vary $\nu$ and show that $J$ is a decreasing function of $\nu$, as expected for pinned chains [\onlinecite{rodh}].

\begin{figure}[!h]
\includegraphics[width=.8\linewidth]{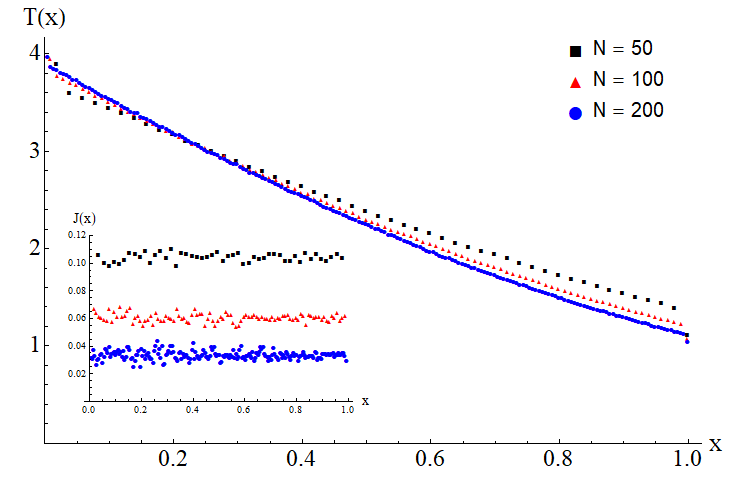}
\caption{Temperature profiles and currents for the Toda chain with quartic pinning ($z = 4$), with all parameters the same as in Fig.~\ref{harmpin_N}. Note that with increasing $N$ the profile approaches a smooth curve between $T_L = 4$ and $T_R = 1$. Insert: Averaged site-by-site current profile. The fact these profiles are flat indicates that we averaged in the NESS.}
\label{quarpin_N}
\end{figure}

We show a marked change in behavior in Fig.~\ref{quarpin_N} for the Toda chain with quartic pinning ($z = 4$). There, as $N$ increases, the temperature profiles approach a smooth curve interpolating between $T_L$ and $T_R$, and the current $J$ approximately satisfies $J \sim N^{-\alpha}$, with $\alpha = 0.88$.
\begin{figure}[!h]
\includegraphics[width=.8\linewidth]{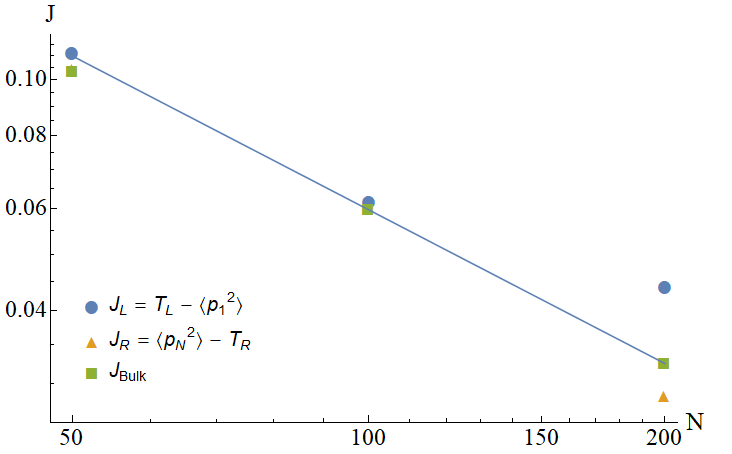}
\caption{Log-log plot of the scaling of $J$ with respect to $N$ for the temperature profiles of Fig.~\ref{quarpin_N}. The straight line was selected to interpolate between $J_{\textrm{Bulk}}$ for $N = 100$ and $N = 200$ with $J_{\textrm{Bulk}} \sim N^{-\alpha}$, $\alpha = 0.88$. We use the bulk current because its variance is reduced by averaging over the chain. Note that $\alpha = 0.88$ differs appreciably from the diffusive scaling exponent of 1, but typically to truly see diffusive behavior one needs to go to larger $N$.}
\label{quarpin_N_2}
\end{figure}

\subsection{Persistent heat currents in the periodic chain}
\label{pheat}
\begin{figure}[!h]
\includegraphics[width=1\linewidth]{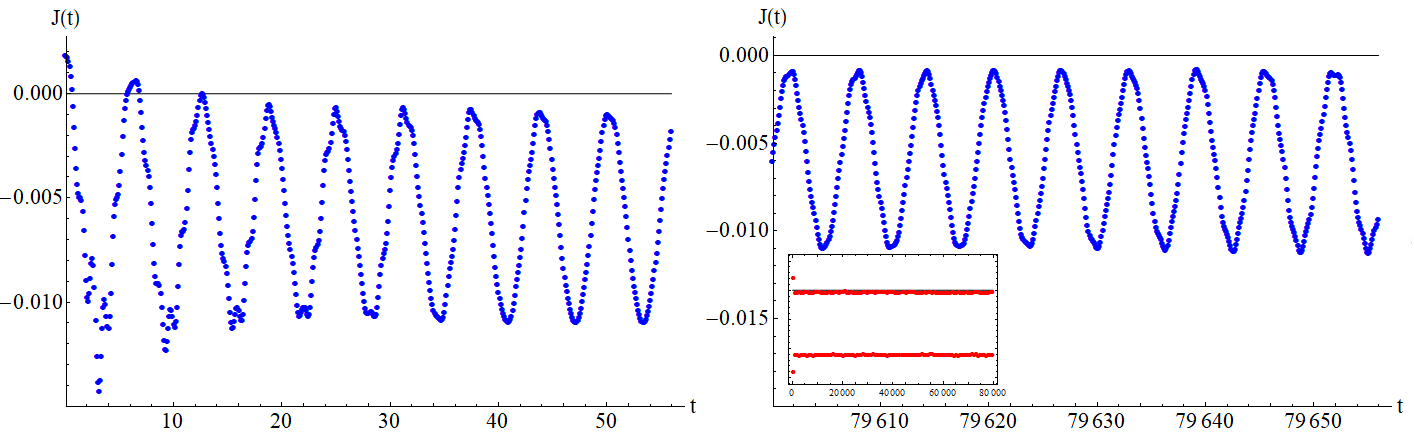}
\caption{Initial and long time behavior of the persistent total current in the Toda chain from \eref{Ham} and \eref{TodaPot} ($a = b = 1$) with harmonic pinning ($z=2$, $\nu = 1$) and periodic boundary conditions, $N = 198$. The dominant frequency of the long time current is very close to the value of the pinning frequency. The dynamics were integrated with a timestep of $dt = 10^{-4}$, and initial condition $q_0(0) = -1$, $p_1(0) = 1$, $q_2(0) = 1$, with all other initial coordinates and momenta zero. The same dynamics with $dt = 10^{-3}$ are nearly identical. Insert: locally averaged maxima and minima of the current for times $t \in[0,8\times10^4]$. Note in Fig.~\ref{diff_current} the decay of the current in the quartic pinning case.}
\label{per_current}
\end{figure}

\begin{figure}[!h]
\includegraphics[width=1\linewidth]{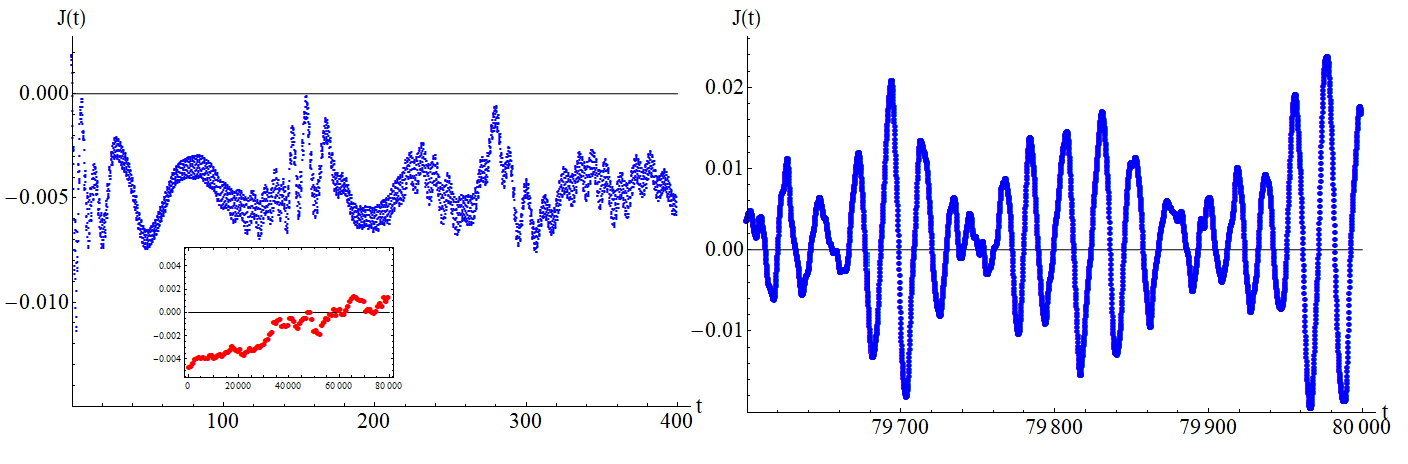}
\caption{Current in the Toda chain with quartic pinning ($z=4$) for the same parameters and initial conditions as Fig.~\ref{per_current}. The current dissipates into noise fluctuating about zero. Insert: Local average of the current for $t \in [0,8\times10^4]$.}
\label{diff_current}
\end{figure}

Consider now the dynamics of the system with Hamiltonian \ref{Ham}, $z=2$, and interaction \ref{TodaPot} with periodic boundary conditions and no external driving. Given the observed ballistic heat transport in the corresponding driven system, one expects any initial heat current to propagate without dissipation in the periodic system. In Fig.~\ref{per_current}, we observe such behavior in a 200 particle system for times long enough for the current to propagate many times around the chain.
Similar nondissipative behavior in the harmonically pinned Toda chain with open boundary was numerically observed in [\onlinecite{ma}]. In that context, the Toda potential \eref{TodaPot} arises as an effective interaction between well-separated solitons of certain solutions of the Gross-Pitaevskii PDE, which models solitons in Bose-Einstein condensates [\onlinecite{weller,coles}]. When the Toda chain was placed in a harmonic trap (which is mathematically identical to pinning each particle in the chain), a Toda soliton was observed to oscillate persistently like the spheres of a Newton's cradle. In diffusive systems, any initial current decays to thermal noise as exemplified in Fig.~\ref{diff_current} when the Toda chain is subject to quartic pinning (z = 4).

\subsection{Poincar\'e sections}
\label{psections}
The observations of Sec.~\ref{btransport} and Sec.~\ref{pheat} for $z=2$ sharply contradict the CAT about the behavior for nonintegrable chains that break momentum conservation. On the other hand, it is possible that the Toda chain with harmonic pinning is in fact integrable. Indeed, the Calogero-Moser Hamiltonian remains integrable when harmonic pinning is added [\onlinecite{cosa}]. While it is highly unlikely that such a simple generalization of a well-known integrable model would have escaped notice for decades, we present dynamical evidence that a higher conservation law exists for $3$ particle Toda chains with harmonic pinning. To do so, we construct Poincar\'e sections of the open chain dynamics, where the end particles are free to move. Each time the particle labeled ``0'' returns to its initial position, we record the momenta of all three particles.

\begin{figure}[!h]
\includegraphics[width=\linewidth]{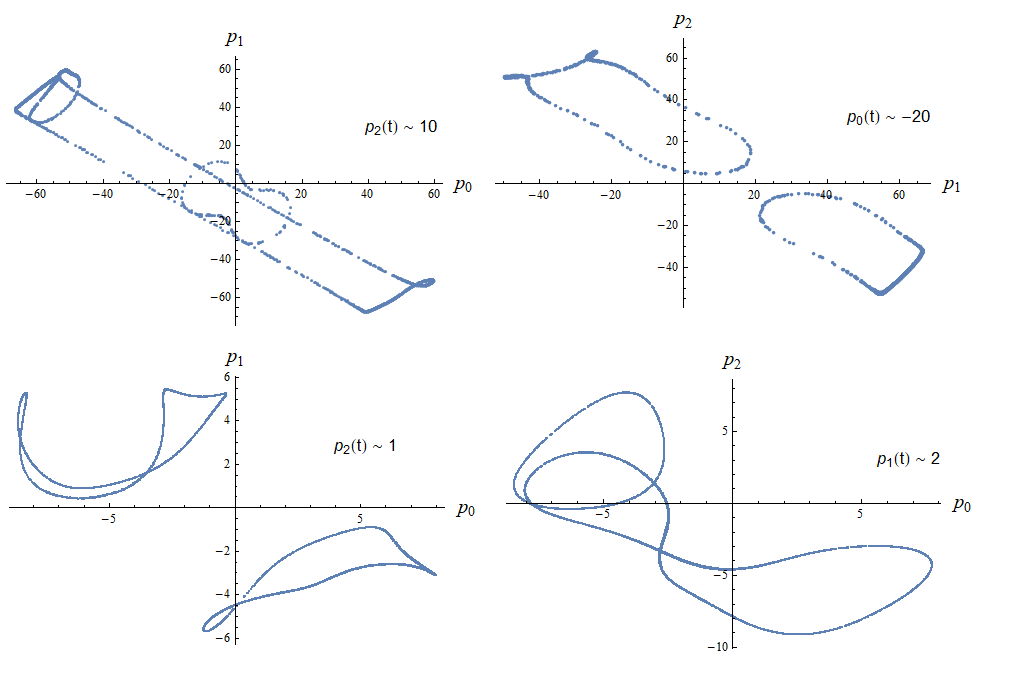}
\caption{Poincar\'e sections for the 3-body open Toda chain ($a = b = 1$) with harmonic pinning ($z=2$, $\nu = 1$). The four images correspond to two different runs (top, bottom) where the momenta $p_j$ were recorded each time $q_0(t) \sim q_0(0)$ (within a tolerance of $\delta = 0.001$). Within each cross section is indicated which momentum is kept fixed and the corresponding value. The tolerance of the fixed momentum is adjusted to allow enough points to make the shape of the curves clear.}
\label{3_sec}
\end{figure}

If the system were nonintegrable, the dynamics of the 3-body case would take place of a 4-dimensional manifold, for there are 6 degrees of freedom corresponding to the positions and momenta, and the conserved Hamiltonian and the conserved term $h_c$ from \eref{c} reduce this number to 6 - 2 = 4. By recording sections when $q_0 = q_0(0)$ and $p_j = p$ for either $j = 0$, $1$ or $2$, we therefore expect to obtain a 4 - 2 = 2 dimensional cross section of the dynamics. For all initial conditions we tested, however, such cross sections are 1-dimensional curves, indicating the presence of an additional conserved quantity. We illustrate this point in Fig.~\ref{3_sec}. If there exists a third conserved quantity that Poisson commutes with $h_c$, the 3-body case is integrable.

We emphasize that the mere existence of initial conditions that constrain the dynamics to lower dimensional manifolds is generic to nonintegrable models. If, however, \emph{all} initial conditions lead to such constrained dynamics, as is the case for Liouville integrability, then it is quite likely that more conserved quantities exist. We tested several random initial conditions for the 3-body case.

Acknowledgements: We are indebted to Abhishek Dhar and Aritra Kundu for providing independent verification of the flat temperature profile. We also thank C\'edric Bernardin, Stefano Olla, Herbert Spohn, Ovidiu Costin, Rodica Costin and Panayotis Kevrekidis for very useful comments. JAS thanks Mitchell Dorrell for his generous computer programming guidance. The work of JLL was supported by AFOSR grant FA9550-16-1-0037. JAS was supported by a Rutgers University Bevier Fellowship.
\FloatBarrier


\begin{thebibliography}{99}
\bibitem{Lepri} S. Lepri, ed. ``Thermal transport in low dimensions.'' Springer International Publishing, 2016.

\bibitem{BLR} F. Bonetto, J. L. Lebowitz and L. Rey-Bellet, ``Fourier's law: a challenge for theorists'', arXiv:math-ph/0002052.

\bibitem{RLL} Z. Rieder, J. L. Lebowitz and E. Lieb, ``Properties of a harmonic crystal in a stationary nonequilibrium state'', J. Math. Phys. \textbf{8}, 1073 (1967).

\bibitem{spohn} H. Spohn. ``Large Scale Dynamics of Interacting Particles.'' Springer-Verlag, Berlin Heidelberg, 1991.

\bibitem{zotos} X. Zotos, ``Ballistic transport in classical and quantum integrable systems'', J. Low Temp. Phys. \textbf{126}, 1185 (2002).

\bibitem{mazur} P. Mazur, ``Non-ergodicity of phase functions in certain systems'', Physics \textbf{43}, 533 (1969).

\bibitem{suzuki} M. Suzuki, ``Ergodicity, constants of motion, and bounds for susceptibilities,'' Physica \textbf{51}, 277 (1971).

\bibitem{kudh} A. Kundu and A. Dhar, ``Equilibrium dynamical correlations in the Toda chain and other integrable models,'' Phys. Rev. E \textbf{94}, 062130 (2016).

\bibitem{toda2} M. Toda, ``Solitons and heat conduction,'' Phys. Scr. \textbf{20}, 424 (1979).

\bibitem{shyo} B. S. Shastry and A. P. Young, ``Dynamics of energy transport in a Toda ring'', Phys. Rev. B \textbf{82}, 104306 (2010).

\bibitem{hatano} T. Hatano, ``Heat conduction in the diatomic Toda lattice revisited,'' Phys. Rev. E \textbf{59}, R1 (1999).

\bibitem{waliha} L. Wang, N. Li and P. H\:anggi, ``Simulation of heat transport in low-dimensional oscillator lattices,'' in [\onlinecite{Lepri}]. 

\bibitem{aolusp} K. Aoki, J. Lukkarinen and H. Spohn, ``Energy transport in weakly anharmonic chains,'' J. Stat. Phys. \textbf{124}, 1105 (2006).

\bibitem{toda} M. Toda, ``Waves in Nonlinear lattice'', Supp. Prog. Theor. Phys. \textbf{45}, 174 (1970).

\bibitem{henon} M. H\'enon, ``Integrals of the Toda lattice'', Phys. Rev. B \textbf{9}, 1921 (1974).

\bibitem{flaschka} H. Flaschka, ``The Toda lattice II: existence of integrals,'' Phys. Rev. B \textbf{9}, 1924 (1974).

\bibitem{wube} J. Wu and M. Berciu, ``Heat transport in quantum spin chains: Relevance of integrability,'' Phys. Rev. B \textbf{83}, 214416 (2011).

\bibitem{AllTil} M. P. Allen and D. J. Tildesley. ``Computer Simulation of Liquids.'' Oxford University Press, New York, 1991.

\bibitem{rodh} D. Roy and A. Dhar, ``Heat transport in ordered harmonic lattices,'' J. Stat. Phys. \textbf{131}, 535 (2008).

\bibitem{ma} M. Ma, R. Navarro, and R. Carretero-Gonz\'alez, ``Solitons riding on solitons and the quantum Newton's cradle,'' Phys. Rev. E \textbf{93}, 022202, (2016).

\bibitem{weller} A. Weller, J. P. Ronzheimer, C. Gross, J. Esteve, M. K. Oberthaler, D. J. Frantzeskakis, G. Theocharis, and P. G. Kevrekidis, ``Experimental Observation of Oscillating and Interacting Wave Dark Solitons,'' Phys. Rev. Lett. \textbf{101}, 130401 (2008).

\bibitem{coles} M. P. Coles, D. E. Pelinovsky, and P. G. Kevrekidis, ``Excited states in the large density limit: a variational approach,'' Nonlinearity \textbf{23}, 1753 (2010).

\bibitem{cosa} E. Corrigan and R. Sasaki, ``Quantum versus classical integrability in Calogero-Moser systems,'' J. Phys. A: Math. Gen. \textbf{35}, 7017 (2002).
\end{thebibliography}
\end{document}